\documentclass[runningheads]{svmult}
 
\usepackage{makeidx}   % allows index generation
\usepackage{graphicx}  % standard LaTeX graphics tool
\usepackage{subeqnar}  % subnumbers individual equations
\usepackage{multicol}  % used for the two-column index
\usepackage{physprbb}  % modified textarea for proceedings,
\makeindex             % used for the subject index
\newcommand{\kms}{km~s$^{-1}$} 

\newcommand{\pns}{PN.S}
\newcommand{\sn}{S/N}                                                                  
\newcommand{\othree}{[O~{\sc III}]}
\newcommand{\Halpha}{H{$\alpha$}}
\newcommand{\microjansky}{$\mu$Jy}

\begin{document}
\bibliographystyle{prsty} 
\title*{Galaxy Dynamics %\protect\newline 
and the PNe Population}
\toctitle{Focusing of a Parallel Beam to Form a Point
\protect\newline in the Particle Deflection Plane}
% allows explicit linebreak for the table of content
%
%
\titlerunning{Galaxy Dynamics and Planetary Nebulae}
% allows abbreviation of title, if the full title is too long
% to fit in the running head
%
\author{Nigel G. Douglas
}
\authorrunning{Nigel Douglas}
% if there are more than two authors,
% please abbreviate author list for running head
%
%
\institute{Kapteyn Astronomical Institute, 
Postbus 800, 9700AV Groningen, The Netherlands}

\maketitle              % typesets the title of the contribution

\begin{abstract} This review \index{abstract} attempts to place the
observations of extragalactic planetary nebulae in the context of
galactic dynamics.  From this point of view only the radial velocities
of the PNe is important. We have built a specialised instrument to
detect PNe  in distant galaxies and measure their radial velocities in
one step. This is explained in some detail, along with classical
techniques for obtaining kinematic information.  The review includes a
vision of possible future developments in the field.

\end{abstract}

\section{Spiral Galaxies}
\begin{figure}[h]
\begin{center}
\includegraphics[width=0.6\textwidth]{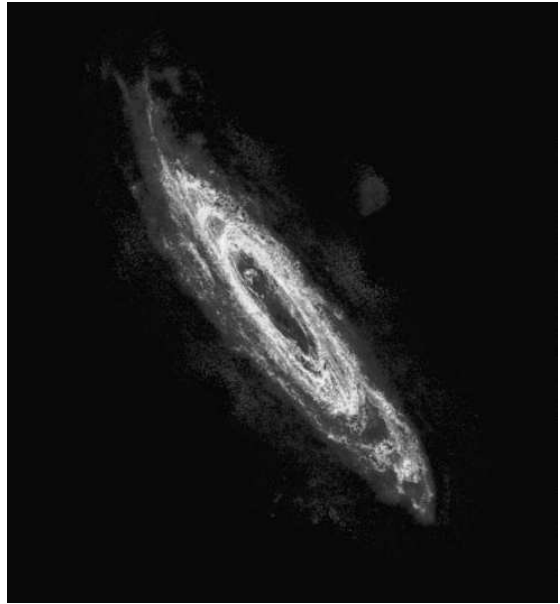}
\end{center}
\caption[]{This radio image of M31 is 
the most detailed view of an H~I disk
ever made, and spans $6^{\rm o}$ on the sky. It can be found in `colour' at
\cite{BraunCWT2002}}
\label{BraunCWT2002}
\end{figure}

Many of the  dynamical properties of galaxies are most easily seen in the
case of spirals.
Their defining property, which is a bright stellar population in coherent rotation, makes it 
relatively easy to determine the size and (apart from a sign ambiguity) the
orientation of the galaxy and its rotation speed as a function of
radius (the ``rotation curve"). It has been found that a large fraction of the neutral 
hydrogen rotates along with the stars, allowing sensitive radio telescopes to map the disk to 
fainter limits (see Fig~\ref{BraunCWT2002} for a state of the art {H~I} image).

 The non-coherent motions of the stars, usually expressed as the standard deviation 
of the three velocity vectors  in  cylindrical coordinates, could be determined, at least for bright galaxies, 
and this gave clues as to the total mass density of the disk. The rotation curve itself provides a measure
of the enclosed mass, and these diagnostics lead to the conclusion that in general considerable dark matter
must be present. Observations of the neutral gas out to large radius from the centre, where the
stellar contribution to the mass density is very small, 
suggest that a dark matter halo becomes
dominant in most cases. 
Note that much of the progress in understanding spiral galaxies is 
possible because the geometry and orientation of the galaxy  is evident,
and to a lesser extent because radio as well as optical observations can
be employed. 

\section{Elliptical Galaxies}

\begin{figure}[h]
\begin{center}
\includegraphics[width=0.8\textwidth]{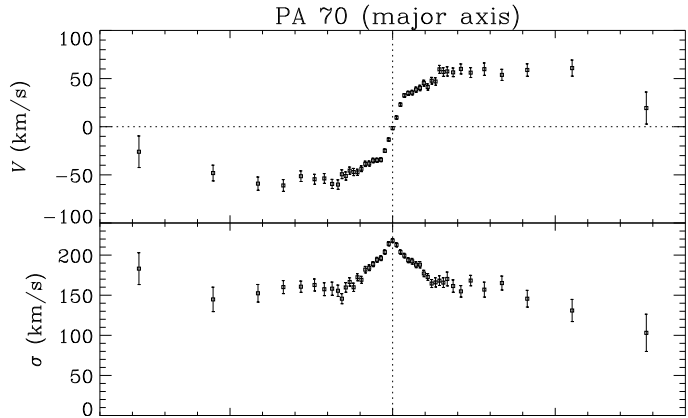}
\end{center}
\caption[]{Major axis rotation in NGC 3379 (from Fig~3a of \cite{StatlerS1999})}
\label{statler1999}
\end{figure}

The situation is quite different in the case of elliptical galaxies  (frequently,
and unfortunately, referred to  as `early-type' galaxies in the
literature). Most of the following remarks apply to the intermediate
type S0 galaxies as well. With their characteristic paucity of neutral 
hydrogen, the kinematics of these galaxies have traditionally been
obtained from  stellar spectroscopy alone, measuring line-of-sight integrals which
are then deconvolved into velocity distributions by template-fitting. 
The effort involved in doing this is considerable.  Moreover, it turns
out that the velocity dispersion is usually comparable to, or larger than, any
systematic rotation which may be
present (see for example Fig~\ref{statler1999}). 
  This  means that, contrary to intuition, rotation alone is not in
general responsible for the observed ellipticity of the galaxy. Much of
the cause must be sought in  incoherent motions, expressed as anisotropy
in the velocity distribution of the stars. The anisotropy is difficult
to determine from the projected velocity information and the orientation
and axis ratio cannot be determined  unambiguously  from the projected
surface brightness. It is difficult even to determine the shape and
orientation of the galaxy - an apparently round, spherically symmetric
galaxy may be elongated along the line-of-sight and even then this axis
can be shorter (oblate galaxy) or longer (prolate) than the tangential
axes. Note, in passing, that it can be shown by simple
order-of-magnitude arguments, that the optical depth  of stars along a
random line of sight through a galaxy is small: galaxies are
transparent. Thus, the  integrated spectrum of the starlight contains
roughly equal contributions from the ``back"  and the ``front" of the
galaxy.

Despite these difficulties, integrated light stellar spectroscopy is
responsible for virtually all of the basic information that we have on
the kinematics of elliptical galaxies. The crucial limitation has turned
out to be that, beyond about 2~$R_e$ (the effective radius $R_e$ being
the mean radius of a contour containing half of the galaxian light) the
surface brightness becomes so low that accurate sky subtraction becomes
impossible. The problem is exacerbated  for relatively nearby galaxies
such as we are talking about here, since the galaxy is so extended
that the sky spectrum generally
cannot be determined in the same exposure, so that that both spatial and
temporal variations degrade the resulting sky subtraction. 
This means that the outer parts
of the galaxies are in effect inaccessible to integrated light spectroscopy.
This is unfortunate as it turns out that the circular velocity 
  curves\footnote{the CVC is derived from the sum of the  velocity and dispersion
  in quadrature.} of nearly all elliptical galaxies are flat, like the rotation
curves of spiral galaxies, out to the last measured point, 
indicating that they too have extended halos. Moreover, some theories of
galaxy formation predict that the outer parts of elliptical galaxies
possess considerable angular momentum. For this and other reasons we
need to be able to explore the kinematics of elliptical galaxies
at large radius.

\section{Kinematic Tracers}
\label{KinematicTracers}
One powerful way to circumvent the limitations of integrated light
spectroscopy is to look for  bright objects in closed orbits in the
vicinity of the galaxy - their motion gives clues to the gravitational
potential of the galaxy and hence to the total enclosed mass. 
Globular clusters (GC) are one such probe, recent examples being 
studies in NGC~4472 and 
Cen~A (NGC~5128) \cite{CoteMC2003,PengFF2004b} respectively.
In these galaxies it turns out that the 
metal-rich and metal-poor GC are kinematically distinct.
In the first, the metal-rich GC show net rotation and the metal-poor ones
did not, while in the second the converse is true. 

Other possible kinematic tracers are satellite 
galaxies and, in nearby galaxies 
such as M31 the kinematics of individual stars can be determined (see the
article by  Ferguson et al.\ elsewhere in the proceedings).

\section{Planetary Nebulae}

During the workshop, spectacular images of PNe have been presented. Even
in the  magellanic clouds, the  complicated morphology and exhilarating
colours of PNe are easily detected. At  greater distances, the PN
envelope reduces to an unresolved point and the colours combine to a 
murky brown. To the extragalactic community, then, the beauty of a PN is
only apparent in its  spectrum, with its characteristic lines of
Hydrogen and `Nebulium.' The value of PNe as kinematic tracers in
external galaxies was demonstrated by Nolthenius and Ford in 1986
\cite{NoltheniusF1986}, combining radial velocity measurements with
dynamical modelling of M32.                                         
Each
PN is seen as a discrete object and it is easy to  measure the radial
velocity of that object by taking the spectrum. At the specific 
wavelength of the \othree\ emission line (5007\AA) 
a single PN can far outshine the local background
of integrated  starlight (the emission lines are inherently narrow and
usually have a measured linewidth of around 20 km/s, or 0.3\AA, as a
result of the expansion velocity of the envelope). PNe are, in principle,
located throughout the body and halo of the galaxy, wherever progenitor
stars are to be found. Unlike `external' probes such as globular
clusters, PNe can be expected to share the orbital properties of the
underlying stellar population, which means that kinematical information
from PNe can be directly combined with stellar kinematics. 
By 1993 several projects had been  completed
and the principal theoretical tools were in place, as summarised in a
seminal paper by  Xiaohui Hui \cite{Hui1993}. Of special note is the
dramatic demonstration, in the case of NGC~3379, of how just a few tens
of PNe can constrain the dynamical models based on integrated light
spectroscopy \cite{CiardulloJD1993}.

Comparisons have been made between PNe and other discrete kinematic
tracers.  It is interesting that in the cases mentioned in
\S\ref{KinematicTracers} the PNe in Cen~A are found to  exhibit
significant rotation \cite{PengFF2004a}, as do the metal-rich GCs in
that galaxy,  while data that we have acquired on NGC~4472
shows that 
the  PNe population has little rotation, 
a different result but again consistent
with the metal-rich GCs. The total gravitating mass of Cen~A
derived from the GCs has also found to be in excellent agreement with
the value derived from  stellar (PN) kinematics \cite{PengFF2004a}.

It is often stated, not least in our own papers \cite{DouglasAF2002},
that PNe are good tracers of the ``old" stellar population. 
I  believe that it is worthwhile keeping in mind that there are
more assumptions in this statement than is usually realised. The main-sequence
lifetime of progenitor stars is {\em not} relevant, at least in the case 
where the galaxy is quiescent with only  very low-level 
continuous star formation. In the steady state the creation rate
of PNe is then {\em independent} of main-sequence lifetime, depends only 
on the initial mass function, and for a typical IMF
is as likely to involve a 
star in the range 2-5$M_{\odot}$  as one in the range 1.5-2$M_{\odot}$. 
With mean ages of 0.6 GYr and 2 Gyr respectively, these populations
may well have different kinematics. 
What presumably tips the scale in favour of the older stars is the {\em lifetime}
of the PNe phase, which is a strong function of progenitor mass \cite{wood}.
On the other hand, PNe formed from high-mass progenitors may be brighter,
and hence preferentially detected in surveys. The point is that, in special
circumstances, one needs to at least question the veracity of the
assumption. The search for possible ``young" PNe in Cen~A (see Peng's paper
elsewhere in this volume) is therefore commendable.

\section{Obtaining Extragalactic PN Velocities}

The traditional technique, pioneered by Hui, Arnaboldi, Ciardullo and others, has been to
identify the PNe by means of a two-filter survey. One is a narrow-band
[O~III] filter centred on the systemic velocity of a target galaxy - both
PNe and foreground stars show up as bright point-like objects - and the other
is a continuum filter in which the stars show up and the PNe do not, 
or are very faint. This enables 
PN candidates to be identified, which  are then 
re-observed using fibre or multi-object spectroscopy (MOS)
to confirm the identification and to measure the radial velocity 
from the \othree\ line.
For sheer volume of effort and persistence over a number of
years, the compilation of 780 confirmed
PNe in Cen~A  must be viewed as one of the exemplary
projects using this technique \cite{PengFF2004a}. 
Amongst other results, the existence 
of PNe out to 15$R_e$ showed that the stellar halo extends that far
(see also Peng's contribution elsewhere in these proceedings).

Another noteworthy approach is the use of a Fabry-Perot interferometer,
which requires no candidate selection and which identifies the
PNe in the entire field (typically a few arcmin) and returns their
velocity (typically with an accuracy of $\sim 30$~\kms) in one
observing run \cite{TremblayMW1995}.

\section{Slitless Spectroscopy and CDI}

Counter-dispersed Imaging (CDI) is a type of slitless spectroscopy in which 
anti-symmetric images are used to determine velocities. The historical antecedents
go back to Fehrenbach (see \cite{DouglasAF2002} and references therein) who realised that  
this procedure not only doubles the effective dispersion, but also diminishes systematic errors.
Once again, Cen~A plays a role in the story, as it was the first extragalactic object  
studied using CDI, by modifying an existing instrument at the Anglo-Australian Telescope.
 The idea was to compare PNe velocities and positions found using CDI 
with those determined earlier by Hui et al.\ \cite{HuiFF1995} using the survey/fibre technique. The results
\cite{DouglasT1999} of the tests were very encouraging: for the 24 PNe found in a certain field
 the astrometry agreed well with the published data except for three cases, for which 
 there was a discrepancy of more than 1~arcsec, and  for precisely these three 
 objects Hui et al.\ had been unable to measure a velocity. 
 In retrospect, it is clear that the fibres missed 
 the object. CDI, on the other hand, is insensitive to astrometric error and
 {\em always} returns a velocity if the PN is
 detected. The first `new' results with CDI on an extragalactic source, using the ISIS
spectrograph at the WHT, were published
 shortly afterwards  \cite{DouglasGK2000}.
 The results of slitless spectroscopy at an 8-m telescope are of course 
dramatic, as exemplified by the detection of 535 PNe in NGC~4697
(see \cite{MendezRK2001} and M\'endez' contribution
elsewhere in this volume). Amongst other results was a distance 
determination of 
around 10.5~Mpc. The relative merits of CDI and other techniques are
discussed in \cite{DouglasAF2002}.

\begin{figure}[h]
\begin{center}
\includegraphics[width=0.8\textwidth]{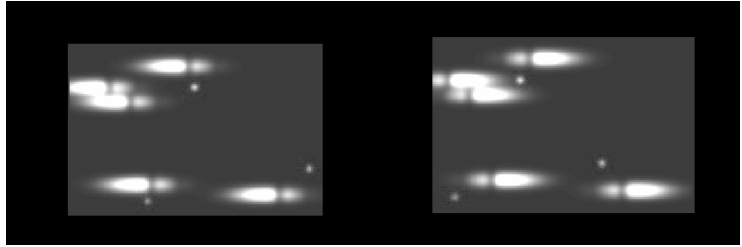}
\end{center}
\caption[]{A simulated pair of  CDI images: the extended trails are stellar spectra, with
an absorption line, convolved by the \othree\ filter profile, while the PNe appear
as dots. The illustration is intended to show how, in principle, 
the relative displacement of the two images of a given PN, 
in combination with suitably calibrated
filter profiles, can return  the radial velocity of the PN. In practice calibration
is done with a special mask.  }
\label{zenit}
\end{figure}

\section{The  PN.Spectrograph}

\begin{figure}[h]
\begin{center}
\end{center}
\includegraphics[width=0.8\textwidth]{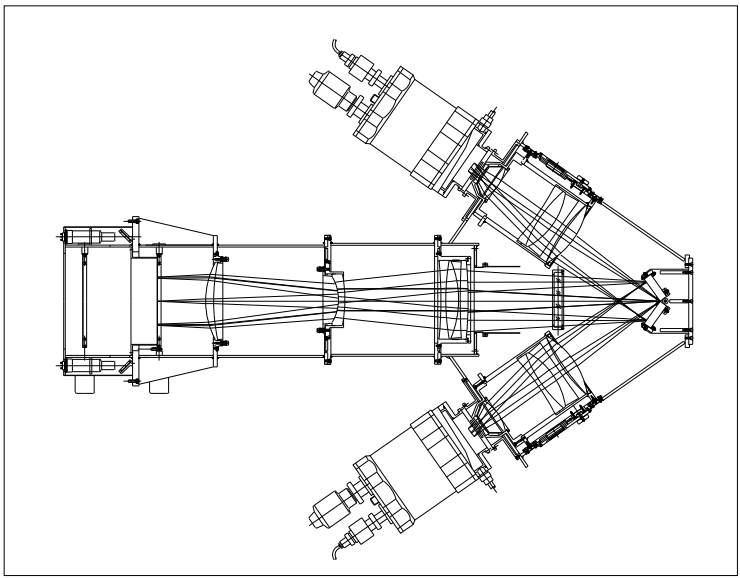}
\caption[]{Schematic of the PN.S - the telescope (not shown)
is to the left.}
\label{pnscdi}
\end{figure}

The PN.Spectrograph (\pns), which  was developed between 1995 and 2001, is intended to optimally
exploit the advantages of CDI described in the previous section. This required 
(a) that the counter-dispersed images should be
obtained simultaneously, not sequentially, (b) that we achieved the highest
possible efficiency in the vicinity of the $\lambda 5007$ \othree\ line,
and (c) that we would  make use of a 4-m class telescope, for which  relatively
large amounts of time are becoming available.
Simultaneity of the  CDI images is important because any systematic
effects (such as seeing) which might  affect the centroiding of the PNe
images, and hence the velocity measurement, are to first order cancelled
out. Moreover one {\em always} has a matched set of data, so that one
clear night guarantees that velocities can be measured,  which was not
the case with our earlier work \cite{DouglasGK2000} and similar programs. 
Figure~\ref{pnscdi} 
shows how we decided to implement simultaneous CDI: a single collimator
illuminates an optical pupil in which two gratings are located, edge to
edge, each diffracting light into its own camera. 
                                                        
\begin{figure}[b]
\begin{center}
\includegraphics[width=0.8\textwidth]{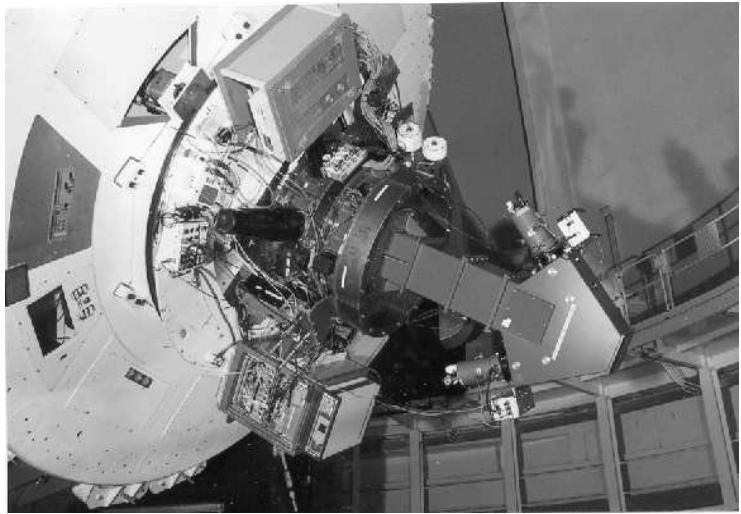}
\end{center}
\caption[]{PN.S at the telescope}
\label{PASPfig9}
\end{figure}

The \pns\ was commissioned 2001 after a  nail-biting delay caused by the
degeneracy of place-names within the Canary Islands - specifically, the
instrument was sent to the wrong island. There were however no further
problems: the  spectrograph worked ``straight from the box" with a
system efficiency of  33\%, including telescope, detector and \othree\
filter. For further technical details see  \cite{DouglasAF2002}.
Figure~\ref{PASPfig9} shows the \pns\ on the telescope. The first major
result with the \pns\ was the confirmation of the paucity of dark matter
in the outer parts of NGC~3379, as suggested by \cite{CiardulloJD1993}
on the basis of 29 PNe. We have now obtained a total of 400 PNe in three
similar galaxies \cite{Romanowsky2003a}. This is part of our core
program, namely the study of a sample of apparently round, 
intermediate-luminosity ellipticals. A side-benefit the \pns\ is that
effectively {\em all} of the observing time is spent in photometric
mode, so that the data should yield good luminosity  functions and thus
distance estimates from the PNLF. On the other hand, the restriction to
just the \othree\ line makes it difficult to recognise interlopers such
as {H~II} regions, which makes the study of later-type galaxies
 more
problematic. To redress this we are in the process of adding a ``piggy-back" H-$\alpha$
camera with about the same sensitivity as the main instrument. As 
has been shown \cite{CiardulloFJ2002}, the relative flux of
 these two lines alone provides a useful discriminant 
against H~II regions .

\section{The Accuracy of  PNe Radial Velocities}

The limitation on the accuracy with which PNe velocities can be measured
is, as usual, set by a combination of the intrinsic linewidth,
instrumental accuracy, and the signal-to-noise which is obtained. For
example, in the  pioneering work of \cite{HuiFF1995}, in which Cen~A
(distance 3 Mpc) was observed with fibre spectroscopy at the ATT, typical
 1$\sigma$ errors were 3--13 \kms, depending on the number of
counts obtained ($\sim 50$). In the case of the \pns\ the velocity
determination of a given PN depends on the accurate centroiding of each
of the two images obtained, as well as, as usual, on the accuracy of the
wavelength calibration. We have established the  centroiding accuracy which
can theoretically be obtained (\cite{DouglasAF2002}, Appendix) and it turns
out that for 
a double detection at the \sn = 5 level 
the separation of the CDI  pairs should be measurable
with an accuracy of about 0.6
pixels, corresponding to 20 \kms\ with our 
modest dispersion of 1.0 pixel \AA$^{-1}$. 
Higher dispersion would help, but CDI does not lend itself to such a
choice for reasons discussed elsewhere \cite{DouglasAF2002}. 
Initial results, in which
we compared measurements of NGC~3379 PNe radial velocities
produced by \cite{CiardulloJD1993}
with  PN.S commissioning data on the same field,  roughly confirm this -
apart from a small systematic trend with velocity, which has yet to be
investigated, the measurements  agreed in absolute heliocentric velocity
to an accuracy of 28 \kms, including  a contribution of $\sim 7$ \kms\
from the earlier measurements.

\section{Future Trends}

Encouraged by the speculative discussion initiated earlier in the
workshop by Quentin Parker, I present some adventurous predictions
with respect to technological advances 
with the potential to push forward the field of galaxy dynamics.

\subsection{Wavelength-tagged Detectors}

It is surprising how few people are aware of the fact 
that detectors are potentially
capable of registering not only the flux at a given position but also
the time of arrival and wavelength of each photon! Some of these
properties were indeed realised by the historic
Image Photon Counting System (IPCS \cite{BoksenbergS1975}), but the last
parameter, wavelength, remains a challenge. If the wavelength of
each photon can be tagged, then we can do away with filters. Any filter
bandpass one wishes to specify can be synthesised {\em after the event}. The
state-of-the-art technology in this area 
is currently the Superconducting Tunnel
Junction  (STJ, \cite{PerrymanFP1999}) which yields a wavelength resolution of about
600\AA\ in the visible region. This is somewhat better than a V-band filter.
Hopefully, with this or perhaps some other technology, we may be able 
to get down to 60\AA\ resolution, at which point one could measure
the \Halpha, \othree, and ``off-band'' flux from every point in the field,
all in one observation, and without having to purchase a single filter
(the \pns\ filters cost \$US~10,000 each). 
This technology would rather radically alter our data-collecting
strategy.

\subsection{Whole-body IFU}

The integral field unit (IFU) was mentioned during this meeting as an
alternative to slit and multi-object spectroscopy in the investigation
of  extragalactic PNe (see articles by Exter, and by Roth, and their
colleagues, elsewhere in this volume). As such, the IFU must be of
potential interest to those of us measuring radial velocities. An  IFU
consists of more-or-less contiguous cluster of fibres, or else a lens
array, covering an area of the field and providing a spectrum of each
position. The field size is in principle limited  only by the `real
estate', in the form of spectrographs and detectors, that one is prepared
to provide. However, this is a non-trivial issue. For galaxy kinematics
with PNe one needs to obtain spectra at large galaxian radius and, even
though the angular size of each aperture could be made larger to 
compensate,  the number
of spectrographs required will still grow quickly.  At large galaxian radius,
where the background is low, it may be remain cost-efficient to survey
for the PNe with two-band images or with slitless spectroscopy. Still,
the idea remains exciting since the concept of `3D
spectroscopy', the registration of a 
complete spectrum at each position on the sky, is hard to beat.
Fortunately, until such monsters  are actually built, there will still be a
place for the \pns.

\subsection{Tangential Velocities$^2$}

\label{vlbi}
\footnotetext[2]{This is work in progress with Jean-Pierre Macquart, until recently
working at the Kapteyn Institute and now at the NRAO, Socorro.}
\addtocounter{footnote}{1}
However well we measure PNe radial velocities, we are still measuring 
only one of the three components of the velocity vector. From an
analytic point of view this  is {\em much} worse than measuring the
radial velocities for only  a third of a sample, say. To illustrate this
consider the extreme but not contrived case of an S0 galaxy viewed pole-on. 
The coherent rotation of the stellar population around the short axis is
entirely in the plane of the sky and thus undetectable. In such
cases, radial-velocity
information alone makes it difficult to 
unambiguously determine anisotropy {\em however many PNe you detect}.
Tangential velocities (i.e. proper motions) would lift this ambiguity completely.
In the example just given it does not much matter whether the proper motions are
determined from the {\em same} objects as those whose radial velocities are known,
but in general it would be best if all three components could be 
determined for a set of test particles. Such a test particle then has 
only one undetermined parameter, its distance along the line-of-sight, whereas if
the tangential and radial velocity information is spread over different particles
then there are two unknowns.

At first glance, the prospect seems totally beyond the realm of the
feasible.  Proper motions at extragalactic distances are tiny: at
the distance of Cen~A (about 3~Mpc) one could wait approximately 
100 years to see
1~milliarcsecond of  proper motion. However, the  next generation of
centimetre-wavelength radio telescopes, of which the Square-kilometre
Array  (SKA) is the design prototype, together with the existing technology
of Very Long Baseline Interferometry
(VLBI) may put this within reach.

The performance goals of the SKA translate to a continuum sensitivity over the
wavelength range of operation, and for a
 one hour integration with 512 MHz bandwidth (roughly equal to the 
internal linewidth of a PN), of 72 nJy beam$^{-1}$. We will see that 
this is easily sufficient to be able to detect 
PNe at extragalactic distances,  but we
ideally also want to be able to detect the {\em same} PNe optically to obtain radial
velocities.   I therefore took the 962 Galactic PNe from the 
Strasbourg-ESO catalogue  \cite{AckerOS1994} for which a reliable 
\othree\ flux measurement was available, and correlated this with the
radio  catalogue of Siodmiak and Tylenda \cite{SiodmiakT2001}. 
This catalogue has 264
entries, of which 232 have measured \othree\ ($\lambda 5007$) fluxes. 
Next, I asked which
of the PNe would potentially be visible at extragalactic distances. 
In optical work,  a PN magnitude\footnote{PN magnitudes are defined
by eqn(2) in the article by Robin Ciardullo elsewhere in this volume;
the apparently arcane zero-point is chosen so that a PN of a given magnitude
appears equally bright as a star of that magnitude seen through a
standard V-band filter.} of $m \sim 28 $  is usually
the  practical limit.
Assuming for simplicity that the galactic PNe are at  a distance of 
10~kpc, I shifted them  to the distance, indicated by D28 in 
table~\ref{vlbi},
at which the \othree\ flux would be reduced to $m = 28$ and calculated
the corresponding 5~GHz flux.  To be conservative  I
rejected those PNe for which the flux at D28 was less than 500~nJy.
This left 47 objects, as listed in the table. No statistical conclusions
can be drawn from this inhomogeneous procedure, but the point to note is
that there exist at least {\em some} PNe whose radio and optical
fluxes would be detectable out to several Mpc. 
The question of whether the observation is feasible in practice is still
to be investigated - 
it may be that, at this flux level, confusion rather than sensitivity may be the main
problem.

How does the astrometric resolution offered by the SKA compare to that
required to observe the transverse motions of planetary nebulae?
We would like to measure the transverse velocity of a PN to within
10\%, which implies that we measure the angular displacement to 
the same accuracy.   Assume that the
typical transverse speed of a planetary nebula is $v\sim 200\,$km/s. 
For a galaxy at a distance $D$ a typical PN then moves an angular
displacement:

\begin{eqnarray}
\Delta \theta =  42 \, \left( \frac{T}{1\,{\rm yr}} \right) \left( \frac{v}{200\,{\rm km/s}} \right) \left( \frac{D}{1\,{\rm Mpc}} \right)^{-1} \quad \mu{\rm as}
\end{eqnarray}
over a time interval $T$.  
If SKA is part of a VLBI network with a 5000\,km baseline we estimate that
at 5~GHz and with an attainable \sn\ of 100 we would obtain a
positional accuracy of $7\,\mu$as. Note that there is no incentive to
improve much on that, since the error in $D$ places a limit of about 10\%
on the
accuracy to which $v$ can be measured. SKA would reach this accuracy
for nearby ($\sim 1$Mpc) galaxies in just a couple of years.
A more serious
concern is the stability of local emission.  The appearance or
disappearance of local hot spots on the ring could offset the emission
centroid significantly.   It would be ironic if the beautiful  internal
structure of PNe, 
 ignored by the galaxy kinematicists, were to 
return in this way to haunt them.

\begin{center}

 \begin{table}[h]
 \caption{Optical and Radio fluxes of some PNe - the columns give
name, measured flux at 5007 and 5~Ghz, D28 (the distance at which the
apparent magnitude at $\lambda$5007 would be $m = 28$) 
and the correspondingly recalculated 5~Ghz flux.  }
\smallskip

\renewcommand{\arraystretch}{1.4}
\setlength\tabcolsep{4pt}  % orig 4pt

\scriptsize
 \begin{tabular}{llrrr |  llrrr}
 \hline\noalign{\smallskip}
  Name      &   Flux      &  5GHz    &  D28    &  5GHz &   Name      &   Flux   &  5GHz    &  D28     &  5GHz\\
            &   5007\AA   &  (mJy)   &  (Mpc)  &  (\microjansky)&             &  5007\AA &  (mJy)   &  (Mpc)   &  (\microjansky)\\

  \noalign{\smallskip}
  \hline
  \noalign{\smallskip}

 000.7+03.2 &     3.90e-13 &       15.6 &        1.4 &         0.80 &  060.5+01.8 &     5.85e-13 &       26.9 &        1.7 &         0.92 \\
 000.7+04.7 &     5.87e-14 &       12.8 &        0.5 &         4.36 &  062.4-00.2 &     6.60e-13 &       17.0 &        1.8 &         0.52 \\
 001.3-01.2 &     1.80e-14 &        9.6 &        0.3 &        10.68 &  064.7+05.0 &     8.40e-12 &      245.0 &        6.5 &         0.58 \\
 001.7+05.7 &     9.12e-13 &       24.4 &        2.1 &         0.53 &  067.9-00.2 &     8.52e-14 &       17.3 &        0.7 &         4.06 \\
 002.8+01.7 &     9.93e-15 &       13.8 &        0.2 &        27.78 &  088.7+04.6 &     4.04e-14 &       15.2 &        0.4 &         7.52 \\
 003.6+03.1 &     1.36e-13 &       23.7 &        0.8 &         3.49 &  093.3-00.9 &     9.38e-13 &       37.3 &        2.2 &         0.80 \\
 005.5+06.1 &     8.98e-14 &       11.8 &        0.7 &         2.63 &  093.5+01.4 &     3.93e-13 &      372.0 &        1.4 &        18.94 \\
 006.2-03.7 &     2.06e-13 &       11.4 &        1.0 &         1.11 &  095.2+00.7 &     1.07e-12 &       59.9 &        2.3 &         1.12 \\
 008.3-01.1 &     3.66e-12 &      163.6 &        4.3 &         0.89 &  096.3+02.3 &     5.24e-13 &       16.9 &        1.6 &         0.65 \\
 011.9+04.2 &     2.51e-12 &       70.5 &        3.5 &         0.56 &  098.2+04.9 &     1.07e-12 &       28.1 &        2.3 &         0.52 \\
 015.9+03.3 &     6.29e-14 &       58.6 &        0.6 &        18.62 &  232.8-04.7 &     1.45e-13 &       26.1 &        0.9 &         3.61 \\
 016.4-01.9 &     1.77e-12 &       78.4 &        3.0 &         0.89 &  235.3-03.9 &     4.02e-13 &       22.2 &        1.4 &         1.10 \\
 020.9-01.1 &     5.33e-13 &      249.0 &        1.6 &         9.34 &  253.9+05.7 &     1.99e-13 &        6.0 &        1.0 &         0.60 \\
 024.1+03.8 &     7.75e-13 &       49.1 &        2.0 &         1.27 &  355.9+02.7 &     4.44e-14 &       21.6 &        0.5 &         9.74 \\
 024.8-02.7 &     4.93e-14 &       13.6 &        0.5 &         5.51 &  356.5-03.9 &     4.47e-14 &       11.1 &        0.5 &         4.96 \\
 027.6+04.2 &     5.00e-13 &       20.9 &        1.6 &         0.84 &  358.3+03.0 &     1.25e-14 &        6.9 &        0.2 &        11.06 \\ 
 034.0+02.2 &     1.21e-13 &       31.0 &        0.8 &         5.10 &  358.5+05.4 &     2.31e-12 &      280.0 &        3.4 &         2.43 \\
 043.1+03.8 &     1.20e-13 &       21.3 &        0.8 &         3.54 &  358.6+01.8 &     1.35e-13 &        6.0 &        0.8 &         0.89 \\
 045.9-01.9 &     1.37e-14 &        9.1 &        0.3 &        13.29 &  358.9+03.2 &     4.67e-15 &       27.3 &        0.2 &       117.01 \\
 048.1+01.1 &     3.29e-13 &       14.2 &        1.3 &         0.86 &  358.9+03.4 &     2.41e-13 &       12.4 &        1.1 &         1.03 \\
 052.9+02.7 &     3.00e-13 &       17.3 &        1.2 &         1.15 &  359.3-03.1 &     8.20e-14 &       11.4 &        0.6 &         2.78 \\
 055.2+02.8 &     4.84e-13 &       23.2 &        1.6 &         0.96 &  359.7-01.8 &     6.83e-13 &       21.5 &        1.8 &         0.63 \\
 056.0+02.0 &     1.14e-13 &       14.6 &        0.8 &         2.57 &  359.8+03.7 &     3.51e-14 &       15.6 &        0.4 &         8.89 \\
 058.9+01.3 &     6.68e-13 &       17.2 &        1.8 &         0.51 &             &              &            &             &             \\

% 23 %24
 
 \hline
  \end{tabular}

%  \end{center}
  \label{Tab1a}
  \end{table}
\end{center}

% How ... displacement 

\section{Acknowledgements}

Thanks to R.~Braun of ASTRON for making available 
Fig~\ref{BraunCWT2002}, to T.S.~Statler of Ohio University
for helpful comments and for making available
Fig~\ref{statler1999}, and to N.~Napolitano for 
useful discussions. We are grateful for the continuing
support of the Isaac Newton Group of telescopes
at La Palma.

\end{document}